\documentclass[aps,twocolumn]{revtex4}
\usepackage{epsf}

\begin{document}
 
\title{Feshbach resonance described by 
       the boson fermion coupling}

\author{T.\ Doma\'nski}
\affiliation{Institute of Physics, M.\ Curie Sk\l odowska University, 
             20-031 Lublin, Poland}
\date{\today}

\begin{abstract}
We consider a possibility to describe the Feshbach resonance in terms 
of the Boson Fermion (BF) model. Using such model we show that after 
a gradual disentangling of the boson from fermion subsystem the 
resonant type scattering between fermions is indeed generated. 
We decouple the subsystems via: (a) the single step, and (b) 
the continuous canonical transformation. With the second one 
we investigate the feedback effects effectively leading to 
the finite amplitude of the scattering strength. We study 
them in detail in: the normal $T>T_{c}$ and superconducting 
$T\leq T_{c}$ states.
\end{abstract}

  
\maketitle

\section{Introduction}
Superconductivity is a quantum state which appears at sufficiently
low temperatures $T\leq T_{c}$ in various systems like e.g.\ metals 
(Hg, Pb), alloys (Nb$_{3}$Sn), copper oxides (La-Sr-Cu-O, Y-Ba-Cu-0), 
exotic compounds (MgB$_2$, UGe$_2$, ZrZn$_2$), the liquid $^{3}$He, 
etc. Depending on a material, $T_{c}$  can range from 10$^{-3}$ to 
more than hundred K and various underlying mechanisms could be 
responsible for superconductivity (like the phonon or magnetically 
mediated attraction between electrons, the BE condensation of tightly 
bound electron pairs, etc). One should add to this list a recently 
obtained superfluidity in the binary mixtures of the trapped alkali 
atoms such as: rubidium \cite{Rb}, potassium \cite{K} and lithium 
\cite{Li}. Transition temperatures $T_{c} \sim 100$ nK became there 
experimentally accessible due to the ``resonance superfluidity''. 
Atoms of different hyperfine configuration are scattered from each 
other with a strength depending on the external magnetic field $B$ 
in a following way $a = a_{0} \left[ 1 - \Delta B/(B-B_{res}) \right]$. 
Near the resonance $B \sim B_{res}$ atoms experience a considerably 
amplified attractive interaction which gives rise to $T_{c}$'s 
comparable with the Fermi temperature $T_{F}$.

From the theoretical point of view such controlled way of 
adjusting the effective interactions between atoms is very 
appealing. It opens new possibilities to explore for instance 
such fundamental problems like a crossover from the weak 
coupling BCS superconductivity to the BE superfluidity of 
tightly bound atom pairs. On a microscopic level, the resonant 
type interactions are however rather delicate to treat (for 
example near the resonance the usual perturbation theory can 
not be applied). So far, the most reasonable way of describing 
such interactions was proposed by Timmermans \cite{Timmermans} 
in terms of the boson fermion (BF) model. This idea was recently 
intensively investigated by the JILA group \cite{Kokkelmans} 
and independently by Griffin and Ohashi \cite{Ohashi}.

Alkali atoms of different hyperfine configurations are 
represented within the BF model via fermion fields which are
characterized by a two labeled index, e.g.\ $\sigma=\uparrow,
\downarrow$. These fermions are assumed to interact with a 
boson field - bosons can be thought of as some bound molecules
made of two atoms. According to the theory \cite{Timmermans}
the resonant type interaction arises when a total energy 
of two colliding fermions matches the energy of boson state, 
the so called Feshbach resonance \cite{Feshbach}.

Using the BF model in the above mentioned context it was 
shown that transition temperature $T_{c}$ can become extremely
high, of the order 0.2 $T_{F}$ for a uniform gas and about 
0.5 $T_{F}$ in an isotropic harmonic trap \cite{Kokkelmans,Ohashi}. 
Since the effective pairing potential modulates from weak to strong 
values the crossover study, along the lines of Nozi\`eres and 
Schmitt-Rink \cite{NSR} was carried out \cite{Kokkelmans,Ohashi} 
clarifying such important issues like the collective modes, etc.

It is worth to recall that the BF model has been known in the 
solid state physics for almost 20 years \cite{Ranninger-85}.
It suits very well for studying superconductivity in systems 
with the arbitrary effective coupling strength. Originally, 
the model was introduced \cite{Ranninger-85} for description 
of the electron-phonon system in a regime between the adiabatic 
and antiadiabatic limits. Later on, also the other possible 
relations were pointed out, e.g.\ to: 
  the periodic Anderson model \cite{Robaszkiewicz-87},
  the extended Hubbard model \cite{Friedberg-94},
  the 2D Hubbard model in its strong interaction 
  limit \cite{Auerbach-02}, and 
  the $t-J$ model \cite{Kochetov-02}.
There is a rich literature discussing properties of the superconducting 
and normal phases described by this model. Many of such properties were 
experimentally observed in the high temperature superconductors (HTSC) 
so the BF model is considered as a candidate for their description.

Since it becomes very popular also for description of the trapped 
alkali atoms, we want to check in this short report if a resonant 
type scattering is really possible to be generated within the BF model. 
We shall prove that indeed, the boson fermion interaction may
give rise to the effective resonant type scattering between fermions. 
This result is obtained by us via: 1) the standard canonical 
transformation and 2) using a continuous canonical transformation 
capable to trace the feedback effects \cite{Domanski-01}. Our study 
thus justifies applicability of the BF model for a description 
of the Feshbach type interactions \cite{Timmermans,Kokkelmans,Ohashi}.

\section{The model}

For simplicity we consider here a model of free fermions coupled 
to the molecular boson field as described by the following 
Hamiltonian \cite{Ranninger-85} 
\begin{eqnarray}
H^{BF} & = & \sum_{{\bf k},\sigma} \left( \varepsilon_{\bf k} 
- \mu \right) c_{{\bf k}\sigma}^{\dagger} c_{{\bf k}\sigma} 
+ \sum_{\bf q} \left( E_{\bf q} - 2\mu \right) 
b_{\bf q}^{\dagger} b_{\bf q} 
\nonumber \\ & + & 
v \sum_{{\bf k},{\bf q}} \left(  b_{\bf q}^{\dagger} 
c_{-{\bf k}+{\bf q}\downarrow}c_{{\bf k}\uparrow} 
+ \mbox{h.c.} \right) \;.
\label{BF}
\end{eqnarray}
neglecting the fermion fermion interactions which eventually 
would be responsible for the {\em background scattering}
\cite{Timmermans,Kokkelmans,Ohashi}. We use the second 
quantization operators $c_{{\bf k}\sigma}^{\dagger}$, 
$c_{{\bf k}\sigma}$ for the fermion state of energy 
$\varepsilon_{\bf k}$ which can exist in two possible 
hyperfine configurations symbolically denoted by 
$\sigma=\uparrow,\downarrow$ and $b_{\bf q}^{\dagger}$, 
$b_{\bf q}$ for the molecular boson state of energy 
$E_{\bf q}$. Fermion and boson subsystems are coupled 
through the isotropic interaction $v$ (in the context of HTSC, 
it should be anisotropic \cite{Micnas-01}). Since a pair of 
fermions can ``dissociate'' into the boson state it implies 
that bosons are carrying the double fermions' charge. 
In order to satisfy the charge conservation we introduce 
the common chemical potential $\mu$ and work in the grand 
canonical ensemble.

It is worthwhile to comment that within the BF model
the mechanism of superconductivity is unconventional  
\cite{Ranninger-85,Robaszkiewicz-87,Friedberg-89,Ranninger-95}. 
Due to interaction $v$ bosons acquire a well established 
effective mass $m_{B}^{-1} \neq 0$, even if at the 
outset they are immobile  ($E_{\bf q}=\Delta_{B}$). At a 
critical temperature $T_{BE}^{B}$ bosons undergo the BE 
condensation and simultaneously this triggers a superconducting 
ordering of the fermion subsystem $T_{sc}^{F} =T_{BE}^{B} 
\equiv T_{c}$ \cite{Kostyrko-96,Domanski-01}.  

Let us point out some of the unusual properties predicted 
theoretically on a basis of this BF scenario which are relevant 
for the HTSC materials but could possibly manifest somehow also
in the mixtures of the trapped alkali atoms. As far as 
superconducting phase is concerned: 
($i$) critical temperatures $T_{c}$ is known to become extremely 
      high because the effective pairing potential [on a level of 
      the mean field theory estimated to be $|V(T)|=\frac{v^{2}}
      {\Delta_{B}-2\mu}\tanh\left(\frac{\Delta_{B}-2\mu}{2k_{B}T} 
      \right)$ \cite{Robaszkiewicz-87}] is very large in the, 
      so called, mixed regime of coexisting fermion and boson 
      particles \cite{Ranninger-85,Robaszkiewicz-87} 
      which occurs when $\mu \rightarrow \Delta_{B}/2$);
($ii$) in the mixed regime, high values of $T_{c}$ are 
      accompanied  by the non-BCS ratio $\Delta_{sc}(0)/
      k_{B}T_{c}>4$ \cite{Robaszkiewicz-87,Ranninger-95};
($iii$) the upper critical field $H_{c2}(T)$ has a characteristic 
      upward curvature $d^{2}H_{c2}(T_c)/dT^{2}>0$ \cite{Hc2_03}.
In the normal phase above $T_{c}$ it was shown that:
($i$) $dc$ resistivity is linear with respect to $T$  
      up to very high temperatures \cite{Eliashberg-87},
($ii$) depending on a doping level and on temperature 
      there is a change of sign of the Hall constant 
      \cite{Geshkenbein-97},
($iii$) in a temperature regime $T^{*} > T > T_{c}$ the 
      pseudogap (partial depletion of the fermion density 
      of states) builds up near the Fermi energy
     \cite{perturbative,DMFT,Domanski-01},
($iv$) the single particle spectrum reveals a clear particle-hole 
      asymmetry \cite{Auerbach-02,Geshkenbein-97,Domanski-03}.

\section{The single step transformation}

For studying effective physics of the BF model it is convenient 
first to apply the standard canonical transformation. We treat 
the boson fermion interaction as a perturbation 
$H_{pert}=v \sum_{{\bf k},{\bf q}} \left(  b_{\bf q}^{\dagger} 
c_{-{\bf k}+{\bf q}\downarrow}c_{{\bf k}\uparrow} + \mbox{h.c.} 
\right)$ and try to eliminate it from (\ref{BF}) via the unitary 
transformation $e^{S}$. Choosing 
\begin{eqnarray}
S & = & \sum_{{\bf k},{\bf q}} \left( \frac{v \; b_{\bf q}^{\dagger} 
c_{-{\bf k}+{\bf q}\downarrow}c_{{\bf k}\uparrow}}{E_{\bf q}-
\varepsilon_{\bf k} - \varepsilon_{-{\bf k}+{\bf q}}} 
- \mbox{h.c.} \right)  ,
\label{S_canon}
\end{eqnarray}
such that $H_{pert}+\left[S,H_{0}\right]=0$, we obtain the 
transformed Hamiltonian $\tilde{H} \equiv e^{S} H e^{-S}$ as
\begin{eqnarray}
\tilde{H} &  = & 
\sum_{{\bf k},\sigma} \left( \tilde{\varepsilon}_{\bf k} 
- \mu \right) c_{{\bf k}\sigma}^{\dagger} c_{{\bf k}\sigma} 
+ \sum_{\bf q} \left( \tilde{E}_{\bf q} - 2\mu \right) 
b_{\bf q}^{\dagger} b_{\bf q} 
\nonumber \\ & + & 
\sum_{{\bf k},{\bf p},{\bf q}} U_{{\bf k},{\bf p},{\bf q}}
c_{{\bf k}\uparrow}^{\dagger} c_{{\bf p}\downarrow}^{\dagger}
c_{{\bf q}\downarrow}c_{{\bf k}+{\bf p}-{\bf q}\uparrow} +
 o(v^{3}) \;,
\label{single_eff}
\end{eqnarray}
where $o(v^{3})$ stands for other terms of the order $v^{3}$.
The renormalized quantities $\tilde{\varepsilon}_{\bf k}$,
$\tilde{E}_{\bf k}$ and $U_{{\bf k},{\bf p},{\bf q}}$ present 
in (\ref{single_eff}) are given by
\begin{eqnarray}
\tilde{\varepsilon}_{\bf k} & = & \varepsilon_{\bf k} + 
v^{2} \sum_{\bf q} \frac{f_{BE}(E_{\bf q})}{\varepsilon_{\bf k} 
+ \varepsilon_{-{\bf k}+{\bf q}}-E_{\bf q}} 
\\
\tilde{E}_{\bf q} & = & E_{\bf q} - v^{2} \sum_{\bf k} \frac{1-
f_{FD}(\varepsilon_{\bf k})-f_{FD}(\varepsilon_{-{\bf k}+{\bf q}})}
{\varepsilon_{\bf k} + \varepsilon_{-{\bf k}+{\bf q}}-E_{\bf q}} 
\\
U_{{\bf k},{\bf p},{\bf q}} & = & \left[ \frac{v^{2}/2}
{\varepsilon_{\bf k} +\varepsilon_{\bf p}-E_{{\bf k}+{\bf p}}} + 
\frac{v^{2}/2} {\varepsilon_{\bf q}+\varepsilon_{{\bf k}+{\bf p}-{\bf q}}
-E_{{\bf k}+{\bf p}}} \right]
\label{eqn6}
\end{eqnarray}
with $f_{BE}$, $f_{FD}$ denoting the Bose-Einstein and Fermi-Dirac
distributions respectively. In the ${\bf q}={\bf p}$ channel thus 
induced interaction between fermions of different $\sigma$ simplifies 
to
\begin{eqnarray}
\frac{v^{2}}{\varepsilon_{\bf k} +\varepsilon_{\bf p}
-E_{{\bf k}+{\bf p}}} \;\; c_{{\bf k}\uparrow}^{\dagger} 
c_{{\bf k}\uparrow}  \;\; c_{{\bf p}\downarrow}^{\dagger} 
c_{{\bf p}\downarrow} 
\label{F_resonance}
\\ \nonumber 
\end{eqnarray}
which explicitly exhibits a divergence. We notice that a scattering 
potential has a resonant type character when total energy of two 
colliding fermions is equal to the energy of boson field 
(Feshbach resonance \cite{Feshbach}).

One may argue that validity of canonical transformation is
limited only to such states which are far from the resonance
because otherwise the operator (\ref{S_canon}) is ill defined. 
Our conclusion about the resonant scattering strength 
(\ref{F_resonance}) may then seem questionable. However, 
in the next section we shall prove that such resonance is 
not an artifact, it indeed exists although somewhat smeared.

\section{Continuous canonical transformation}

Instead of a single step transformation we will now 
disentangle the fermion from boson subsystems applying
a sequence of infinitesimal transformations. The main 
idea behind is to proceed with a continuous canonical
transformation $S(l)$ (where $l$ denotes some formal
{\em flow parameter} which can vary between 0 and any 
other value) until a given structure of the Hamiltonian
$H(l)=e^{S(l)}He^{-S(l)}$ is obtained. A virtue
of this method is that one can freely manipulate with 
$S(l)$ in order to get a constrained structure of $H(l)$.

In this work we follow the algorithm outlined by Wegner 
\cite{Wegner-94} for constructing the generating operator 
$\eta(l) \equiv dS(l)/dl$ of the continuous transformation.
An arbitrary Hamiltonian $H(l)=H_{0}(l)+H_{pert}(l)$ can 
be reduced to a semidiagonal structure by choosing $\eta(l) 
= [H_{0}(l),H_{pert}(l)]$ which assures that $\lim_{l\rightarrow
\infty}H_{pert}(l)=0$. During the transformation $H_{0}(l)$ part 
does evolve too, its $l$-dependent parameters have to be deduced  
from the general flow equation $dH(l)/dl=[\eta(l),H(l)]$ 
\cite{Wegner-94}. In practice, this flow equation can never 
be identically satisfied because the higher and higher order 
interactions (not present in $H_{0}$) are generated from the 
commutator. Flow equation is often approximated by truncating 
the higher order interactions in a spirit of the perturbation 
theory, however the nonperturbative methods are in principle 
possible too \cite{Kehrein-99}.

In the previous work \cite{Domanski-01} we have already 
formulated the continuous canonical transformation for 
the BF model. The corresponding flow equations (16-21) 
of Ref.\ \cite{Domanski-01} were derived within a perturbational
estimation up to order $v^{3}$. In this short report we want 
to study in some more detail the resonant-type interactions 
induced between fermions in the effective Hamiltonian 
$\tilde{H}=H(l\rightarrow\infty)$.

In a course of transformation the interactions between fermions 
evolve from zero, at $l=0$, to some effective value 
$\tilde{U}_{{\bf k},{\bf p},{\bf q}}$ at $l=\infty$ when boson 
and fermion subsystems are finally decoupled from each other. 
{\em Flow} of the potential is governed by the equation 
\cite{Domanski-01}
\begin{eqnarray}
\frac{dU_{{\bf k},{\bf p},{\bf q}}(l)}{dl} = 
\left[ \alpha_{{\bf k},{\bf p}}(l) + 
\alpha_{{\bf q},{\bf k}+{\bf p}-{\bf q}}(l) \right] 
v_{{\bf k},{\bf p}}(l)v_{{\bf q},{\bf k}+{\bf p}-{\bf q}}(l) 
\nonumber \\ 
\label{interflow} 
\end{eqnarray}
where $\alpha_{{\bf k},{\bf p}}(l)=\varepsilon_{\bf k}(l)+
\varepsilon_{\bf p}(l)-E_{{\bf k}+{\bf p}}(l)$ and the 
introduced momentum dependence of the boson-fermion coupling 
$v_{{\bf k},{\bf p}}(l)b_{\bf q}^{\dagger} c_{-{\bf k}+{\bf q}
\downarrow}c_{{\bf k}\uparrow}$ arises from $dH/dl$.
Equation (\ref{interflow}) is convoluted with the following 
ones \cite{Domanski-01}
\begin{eqnarray}
\frac{dv_{{\bf k},{\bf p}}(l)}{dl} & = & 
- \alpha^{2}_{{\bf k},{\bf p}}(l) v_{{\bf k},{\bf p}}(l) \;,
\label{hybrflow} \\
\frac{d\varepsilon_{\bf k}(l)}{dl} & = & 2\sum_{\bf p} 
\alpha_{{\bf k},{\bf p}}(l) v^{2}_{{\bf k},{\bf p}}(l) 
f_{BE}(E_{{\bf k}+{\bf p}}(l)) \;,
\label{epsflow} \\
\frac{dE_{\bf q}(l)}{dl} & = & - 2\sum_{\bf k}  \alpha_{{\bf q}-{\bf k},
{\bf k}}(l) v^{2}_{{\bf q}-{\bf k},{\bf k}}(l) 
\nonumber \\ & & \times
\left[ 1 -2 f_{FD}(\varepsilon_{{\bf k}-{\bf q}}(l)) \right] .
\label{Ekflow}  
\end{eqnarray}
These four $l$-dependent quantities $v_{{\bf k},{\bf p}}(l)$, 
$\varepsilon_{\bf k}(l)$, $E_{\bf q}(l)$, $U_{{\bf k},{\bf p},{\bf q}}(l)$ 
should be determined simultaneously. Mathematically it is a tremendous 
task, nevertheless we shall estimate $\tilde{U}$ either in an approximate 
way or numerically and, in one special case, exactly.

\subsection{Exact solution at $T<T_{c}$}

Some exact statements can be done for the superconducting/superfluid
phase of the BF model. For temperatures $T$ smaller than a critical 
$T_{c}$ there exists a finite fraction of condensed bosons 
$n_{0}^{B}=\left< b_{\bf q}^{\dagger}b_{\bf q} \right>_{{\bf q}={\bf 0}}$ 
and chemical potential is then located at the lowest boson
energy $\mu(T)=E_{\bf 0}/2$. The flow equation (\ref{epsflow}) 
$d\varepsilon_{\bf k}(l)/dl \simeq 4 \xi_{\bf k}(l) n_{0}^{B} 
v^{2}_{{\bf k},-{\bf k}}(l)$ combined with (\ref{hybrflow}) 
$dv_{{\bf k},-{\bf k}}(l)/dl=-4\xi^{2}_{\bf k}(l) v_{{\bf k},-{\bf k}}(l)$
lead to the following invariance $\xi_{\bf k}(l)^{2}+v_{{\bf k},-{\bf k}}^{2}(l)
=const$, where $\xi_{\bf k} \equiv \varepsilon_{\bf k}(l)-\mu$.
In consequence, for the $l\rightarrow \infty$ limit, we obtain  
the effective Bogolubov type spectrum $|\xi_{\bf k}(\infty)|=
\sqrt{ \xi^{2}_{\bf k} + n_{0}^{B} v^{2}}$ [see Eqn.\ (52) of the 
Ref.\ \cite{Domanski-01}].

Using the flow equation (\ref{interflow}) we find that,
in the ${\bf p}={\bf q}=-{\bf k}$ channel, potential 
$U_{{\bf k},{\bf p},{\bf q}}(l)$ is given by
\begin{eqnarray}
\frac{dU_{{\bf k},-{\bf k},-{\bf k}}(l)}{dl} =  4
\xi_{\bf k}(l) v_{{\bf k},-{\bf k}}^{2}(l) =
 \frac{-1}{2 \xi_{\bf k}(l)} \; \frac{dv_{{\bf k},
-{\bf k}}^{2}(l)}{dl}
\label{eqn12}
\end{eqnarray}
due to $dv^{2}_{{\bf k},-{\bf k}}(l)/dl=-8\xi^{2}_{\bf k}(l) 
v^{2}_{{\bf k},-{\bf k}}(l)$. After integration by parts we 
get from (\ref{eqn12}) 
\begin{eqnarray}
\left. \xi_{\bf k}(l)U_{{\bf k},-{\bf k},-{\bf k}}(l) \right| 
_{0}^{\infty} -\int_{0}^{\infty} \frac{d\xi_{\bf k}(l)}{dl}
U_{{\bf k},-{\bf k},-{\bf k}}(l) = \frac{v^{2}}{2}
\end{eqnarray}
because $v_{{\bf k},-{\bf k}}(\infty)=0$. Since according 
to (\ref{epsflow}) we have $d\xi_{\bf k}/dl \propto v^{2}$ 
and from our previous estimation \cite{Domanski-01} also 
$U_{{\bf k},{\bf p},{\bf q}} \propto v^{2}$  we finally 
conclude
\begin{eqnarray}
U_{{\bf k},-{\bf k},-{\bf k}}(\infty) = 
\frac{v^{2} + o(v^{4})}{ 2\mbox{sign}\left(\xi_{\bf k}\right)
\sqrt{ \xi^{2}_{\bf k} + n_{0}^{B} v^{2}}} \;.
\label{F_sc}
\\ \nonumber 
\end{eqnarray}
This equation shows that two colliding electrons with total
energy $\varepsilon_{\bf k} + \varepsilon_{-{\bf k}}=
E_{{\bf q}={\bf 0}}$ have a resonant-like scattering strength
(remember that in this case $E_{\bf 0}=2\mu$ and due to symmetry 
$\varepsilon_{-{\bf k}}=\varepsilon_{\bf k}$). Amplitude of the 
resonance is now finite and is controlled by the superconducting 
gap $\Delta_{sc}(T)=v\sqrt{n_{0}^{B}(T)}$.

\subsection{Approximate solution for $T \geq T_{c}$}

Effective interaction between fermions can be calculated 
approximately, for instance iteratively. In the first step 
we can neglect $l$-dependence of $\alpha_{{\bf k},{\bf p}}(l)$ 
in the flow equations (\ref{interflow}) and (\ref{hybrflow}) 
because fermion $\varepsilon_{\bf k}(l) \simeq \varepsilon_{\bf k}$ 
and boson energies $E_{\bf q}(l) \simeq E_{\bf q}$ are rather weakly 
renormalized during the transformation \cite{Domanski-01}.

With an approximation $\alpha_{{\bf k},{\bf p}}(l) \simeq 
\alpha_{{\bf k},{\bf p}}(l=0)$ we can easily obtain $v_{{\bf k},
{\bf p}}(l)=v \; \mbox{exp} [-\left( \varepsilon_{\bf k}+
\varepsilon_{\bf p}-E_{{\bf k}+{\bf p}}\right)^{2}l]$. 
When further substituted to equation (\ref{interflow}) 
we get the following asymptotic value
${U}_{{\bf k},{\bf p},{\bf q}}(l \rightarrow \infty)$
\begin{eqnarray}
\frac{v^{2}\left( \varepsilon_{\bf k}+\varepsilon_{\bf p}
 + \varepsilon_{\bf q}+\varepsilon_{{\bf k}
+{\bf p}-{\bf q}}- 2 E_{{\bf k}+{\bf p}} \right)}
{\left( \varepsilon_{\bf k}+\varepsilon_{\bf p}
- E_{{\bf k}+{\bf p}} \right)^{2} +
\left( \varepsilon_{\bf q}+\varepsilon_{{\bf k}+{\bf p}-{\bf q}}
- E_{{\bf k}+{\bf p}} \right)^{2}
} 
\end{eqnarray}
which is less divergent than (\ref{eqn6}). However, in the
${\bf q}={\bf p}$ channel it again reduces to the resonant
type potential given in (\ref{F_resonance}). Divergence at 
$\varepsilon_{\bf k}+\varepsilon_{\bf p}=E_{{\bf k}+{\bf p}}$ 
occurs here because the flow equations (\ref{interflow}-
\ref{Ekflow}) were not investigated fully selfconsistently.

In order to check, whether a selfconsistent treatment 
does not lead to any divergence we solved equations 
(\ref{interflow}-\ref{Ekflow}) numerically by the Runge 
Kutta method. At the starting point $l=0$,  we used 
the parabolic dispersions $\varepsilon_{\bf k}=(\hbar 
{\bf k})^{2}/2m_{e}$,  $E_{\bf q}=\Delta_{B}+(\hbar 
{\bf q})^{2}/2m_{b}$ and introduced the ultraviolet 
cutoff $\Lambda \gg \mu(T)$. 

We obtained the effective potential ${U}_{{\bf k},{\bf p},{\bf q}}
(l \rightarrow \infty)$ for several temperatures $T$ using a fixed 
total charge concentration $\sum_{{\bf k}\sigma} \left< c_{{\bf k}
\sigma}^{\dagger}c_{{\bf k}\sigma} \right> + 2 \sum_{\bf q} 
\left< b_{\bf q}^{\dagger}b_{\bf q} \right> =const$. 
Temperature had rather a negligible influence on a magnitude 
of the fermion fermion interaction for $T>T_{c}$. 
Using a given $\Delta_{B}$ value we adjusted the total 
charge concentration so, that number of bosons $N_{B}=\sum_{\bf q}f_{BE}
(E_{\bf q})$ and fermions $N_{F}=\sum_{\bf k}f_{FD}(\varepsilon_{\bf k})$ 
were comparable. Chemical potential was located very close to the bottom 
of boson states $\mu(T) \simeq \Delta_{B}/2$. Figure 1 illustrates the
behavior of $U_{{\bf k},{\bf p},{\bf p}}(\infty)$. Again, we notice
an appearance of the resonant type interaction for $\varepsilon_{\bf k}+
\varepsilon_{\bf p}=\Delta_{B}$ but, in distinction from 
(\ref{F_resonance}), amplitude of the resonance is finite.

\begin{figure}
\epsfxsize=6.5cm
\centerline{\epsffile{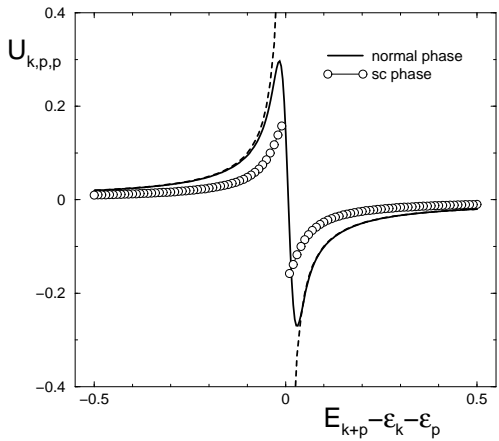}}
\caption{Potential of the effective interaction between fermions 
${U}_{{\bf k},{\bf p},{\bf p}} c_{{\bf k}\uparrow}^{\dagger} 
c_{{\bf p}\downarrow}^{\dagger} c_{{\bf p}\downarrow} c_{{\bf k} 
\uparrow}$ obtained: (i) from the single step transformation 
expressed by the equation (\ref{F_resonance}) (dotted line), and 
(ii) from the continuous canonical transformation. Solid line 
refers to the normal phase $T>T_c$ and circles to the superconducting 
state $T=0<T_c$ respectively. The BF model parameters are: 
$m_{B}=2m_{F}$ and $v=0.1$, $\Delta_{B} = 0.5$ in units
of the Fermi energy.}
\end{figure}

\section{Conclusions}

In summary, we analyzed the resonant type interactions between 
fermions (experimentally achievable for example in the binary 
mixtures of $^{85}$Rb \cite{Rb}, $^{40}$K \cite{K} or $^{6}$Li 
atoms \cite{Li} in a presence of the magnetic field) using 
the microscopic boson fermion (BF) model. Within the BF scenario, 
fermions are coupled to the boson field (the long lived molecules 
composed of two atoms). This coupling is responsible for 
the many body effects and, in particular at $T_{c}$ it drives 
a system to the superconducting/superfluid ordering 
\cite{Ranninger-85}. When approaching the critical 
temperature from above, several precursor features can be observed 
which show that the pairing and the long range pair coherence 
may occur in this model at different temperatures $T^{*}$ and 
$T_{c}\leq T^{*}$, respectively.

We determined the effective fermion fermion interaction induced
by elimination of the boson fermion coupling via the canonical 
transformation. Boson and fermion subsystems were disentangled 
(i) by the single step transformation, and (ii) through a continuous 
sequence of transformations taking account of the feedback effects
\cite{Domanski-01}. In both methods we obtained the effective resonant 
type interactions between fermions. At temperatures $T>T_{c}$ the
resonance appeared to be more sharp (although of finite amplitude), 
while for $T < T_{c}$ its amplitude reduced to a magnitude of the 
superconducting gap. 

The resonant scattering seems to be a robust feature of the mixed 
boson-fermion system. It has an impact on several physical properties, 
for instance, it is responsible for the particle-hole asymmetric 
spectrum \cite{Domanski-03}. Other aspect which is important
in a context of the soft matter physics is that the microscopic 
BF model can serve as a convenient tool for description of 
the resonant type interactions $a = a_{0} \left[ 1 - \Delta B/
(B-B_{res}) \right]$ \cite{Timmermans,Kokkelmans,Ohashi}. 
Many body effects can there be studied in a secure way (though 
still far from trivial) without a necessity to deal with any 
divergences.

\vspace{0.5cm}
Author kindly acknowledges valuable discussions with
Profs J.\ Ranninger, F.\ Wegner and K.I.\ Wysoki\'nski.

\end{document}